# Untraceable Email Cluster Bombs:
# On Agent-Based Distributed Denial of Service

Markus Jakobsson
RSA Security
mjakobsson@rsasecurity.com

Filippo Menczer
The University of Iowa
filippo-menczer@uiowa.edu

## Abstract

We uncover a vulnerability that allows for an attacker to perform an email-based attack on selected victims, using only standard scripts and agents. What differentiates the attack we describe from other, already known forms of distributed denial of service (DDoS) attacks is that an attacker does not need to infiltrate the network in *any manner* — as is normally required to launch a DDoS attack. Thus, we see this type of attack as a *poor man's DDoS*. Not only is the attack easy to mount, but it is also almost impossible to trace back to the perpetrator. Along with descriptions of our attack, we demonstrate its destructive potential with (limited and contained) experimental results. We illustrate the potential impact of our attack by describing how an attacker can disable an email account by flooding its inbox; block competition during on-line auctions; harm competitors with an on-line presence; disrupt phone service to a given victim; cheat in SMS-based games; disconnect mobile corporate leaders from their networks; and disrupt electronic elections. Finally, we propose a set of countermeasures that are light-weight, do not require modifications to the infrastructure, and can be deployed in a gradual manner.

**Keywords:** Distributed Denial of Service, Email, SMS, Web Forms, Agents

## 1 Introduction

The competitive advantage of most industrialized nations depends on a well-oiled and reliable infrastructure, much of which depends on the Internet to some extent. We show how one very simple tool can be abused to bring down selected sites, and argue how this in turn — if cunningly performed — can do temporary but serious damage to a given target. Here, the target may be a person, business or institution relying on the Internet or the telephone network for its day to day activities, but may also be more indirectly dependent on the attacked infrastructure.

When assessing the damage a potential attack can inflict, it is important to recognize that attacks may carry a substantial cost to society even if they do not obliterate their targets — in particular if repeatedly perpetrated, which becomes easier if the attacks are difficult to trace back to their perpetrators. Furthermore, one should not only take the direct costs into account, but also indirect costs, namely those associated with not being able to rely on the infrastructure.

When considering the (in)stability of our infrastructure, it is also crucial to understand that the real target of an attack may be a secondary and indirect one, whose relation to the site being brought down may not be evident until the attack takes place. Therefore, the target may not be the least prepared for an attack of the type it would suffer, making the blow even harder. For example, if voters are allowed to cast votes using home computers or phones (as in recent trials in Britain [9]), then an attack on *some* voters or servers may invalidate the *entire* election, requiring all voters to cast their votes again — for fairness, this would include even those who used traditional means in the first place. Other potential examples of secondary damage include the general mobile phone system, the infrastructure for delivery of electricity from power plants to consumers, and the traffic-balancing of the Interstate highway system, given that these allow for load balancing via the Internet in many places.

**Approach.** The attack involves Web crawling agents that, posing as the victim, fill forms on a large set of third party Web sites (the "launch pads") causing them to send emails or SMSs to the victim, or have phone calls placed. The launch pads do not intend to do any damage — they are merely tools in the hands of the attacker.

Our attack takes advantage of the absence in the current infrastructure of a (non-interactive) technique for verifying that the submitted email address or phone number corresponds to the user who fills in the form. This allows an automated attacker to enter a victim's email or number in a tremendous number of forms, causing a *huge* volume of messages to be directed to the victim's mailbox. Depending on the quantity of generated messages, this may cost the victim anything from lost time (sorting out what mes-sages to delete); to lost messages (if the mailbox fills up, causing the Internet Service Provider (ISP) to bounce legitimate emails); to a crash or other unavailability of some of the victim's or ISP's machines.

We present experimental data indicating the ease of mounting the attack; the time it takes to mount attacks of certain sizes; and the time it takes for a certain quantity of email to be generated by attacks of various sizes.

**Potential victims.** Our attack is applicable both to "conventional" computers and to mobile devices, such as cellular phones, PDAs, and other messaging devices. It is easy to see that cellular phones with text messaging can be attacked in the same way as normal email accounts. Not only does this generate network congestion and unwanted costs, but it also causes the text messaging feature of a mobile phone to be disabled once memory is filled up. According to a quick test of ours, the memory of a common cell phone model fills up after around 80 messages — an attack we performed in a few seconds. We note that an attacker would not have to know what cell phone numbers are in use in order to mount a general attack on the *service provider* — he can simply attack large quantities of numbers at random, many of which will be actual numbers given the high density of used numbers. This type of attack would allow an attacker to stop medical doctors from being paged; inconvenience everyday users of SMS; and cheat in location based games such as Vodafone's *BotFighters*.[1] Moreover, an attacker can target all email accounts with names likely to correspond to a given corporate leader and thereby render her mobile device unable to receive meaningful messages. This could be done for the corporate domain as well as for all common providers of email and mobile connectivity.[2]

The common telephony infrastructure (both mobile and wired) can be attacked in an analogous manner: by agents entering a victim's *phone number* in numerous forms. If the remaining entered information is not consistent or accurate, this may result in a representative of the corresponding company placing a phone call to straighten things out, possibly after trying to send one or more messages to the email address entered in the form. Given the higher cost of placing a phone call — compared to sending an email — many companies prefer responding by email, which is likely to require a larger number of forms to be filled in by an attacker, in order to cause a comparable call frequency. On the other hand, phone calls being more disruptive than email messages, the impact of the attack types may be comparable for a given attack size.

**Defenses.** What complicates the design of countermeasures is the fact that there is nothing *per se* that distinguishes a malicious request for information from a desired request in the eyes of the launch pad site, making the latter oblivious to the fact that it is being used in an attack. This also makes legislation against unwanted emails, SMSs and phone calls [8] a meaningless deterrent: without the appropriate technical mechanisms to distinguish valid requests from malicious ones, how could a site be held liable when used as a launch pad? To further aggravate the issues, and given that our attack is a type of DDoS attack, it will not be possible for the victim (or nodes acting on its behalf) to filter out high-volume traffic emanating from a suspect IP address, even if we ignore the practical problems associated with spoofing of such addresses.

The standard defense against impersonation of users is not useful to avoid the generation of network traffic. In particular, some sites attempt to establish that a request emanated with a given user by sending the user an email to which he is to respond in order to complete the registration or request. However, as far as our email-based attack is concerned, it makes little difference whether the emails sent to a victim are responses to requests, or simply emails demanding an acknowledgement.

While it may appear that the simplicity and generality of the attack would make it difficult to defend against, this is fortunately not the case. We propose (1) simple extensions of known techniques whereby well-intentioned Web sites can protect themselves from being exploited as launch pads for our attack, and (2) a set of heuristic techniques whereby users can protect themselves against becoming victims. Our countermeasures are light-weight and simple, require no modifications of the communication infrastructure, and can be deployed gradually.

**Outline:** We begin by describing how an attack can be mounted using standard agent-based techniques (section 2), noting how easy it is for an attacker to remain untraceable. We then present experimental data supporting the strength of the attack (section 3), followed by a brief survey of possible targets of attack (section 4). We then propose some techniques to secure potential launch pads and targets of different kinds (section 5), noting that not all targets can use the same defense techniques. We finally discuss related work and some open problems related to defense against agent-mounted attacks (section 6).

---

[1] This game (www.botfighters.com) is based on receiving and sending SMSs. Since it does not require any special software to be downloaded, it also cannot lock out messages from places other than the game center. This allows users with knowledge of another player's phone number to mount a denial of service attack, efficiently paralyzing the victim.

[2] If the names of the victims are not known, an attacker can mount a dictionary attack in combination with the DDoS attack we describe.

## 2 The Attack

### 2.1 Description of Vulnerability

Many sites allow a visitor to request information or subscribe to a newsletter. A user initiates a request by entering her contact information in a form, possibly along with additional information. Figure 1 shows a typical form.

Our attack takes advantage of the fact that, in the current Web infrastructure (e.g., HTTP protocol), there is no way to verify that the information a user enters corresponds to the true identity or address of the user. Thus it is possible to request information on behalf of another party. Agents — or automated scripts acting as users — allow this to be performed on a large scale, thereby transforming the illegitimate requests from a poor practical joke to an attack able of bringing down the victim's site.

### 2.2 Finding the Victim

In many instances, the attacker may know the email address or phone number of the victim, or may be able to extract it from postings to newsgroups, replies in an auction setting, etc. In other cases, the address may be unknown. If the attacker wishes to target the corporate leaders of a given company, he has to determine what their likely addresses are, which typically are limited to a few combinations of first and last names. In order to target mobile devices, such as Blackberries, the attacker would also target the appropriate wireless service providers, again targeting all names that match the victim(s). In order to target a service provider, a massive attack of this type is also possible. To wreak havoc in an electronic election in which users are allowed to use their own computers and wireless devices, it suffices to target a few voters, who will later complain that they were locked out. It is even possible for an attacker to block his own device (stopping *himself* from voting) in order to later be able to lodge a complaint and have the election results questioned.

### 2.3 Phase I: Harvesting Suitable Forms

Many Web sites use forms to execute scripts that will collect one or more email addresses and add them to one or more lists. There are many legitimate ways in which the collected emails can be used: mailing lists for newsletters, alert services, postcards, sending articles or pages to friends, etc. There are less legitimate uses as well, for example many sites collect emails by advertising freebies of various sorts, and then sell the email lists to spammers as "opt-in" requests.

One way for an attacker to automatically locate and collect forms to be used as launch pads is by employing a topic-driven crawler [6, 7]. Such a software searches the

```
base = (free email newsletter);
list = (alert subscribe opt-in list spam
    porn contest prize stuff travel ezine
    market stock joke sign verify money
    erotic sex god christ penis viagra age
    notify news recipe gratis libre livre);
foreach set = subset(list) {
    query(base plus(set) minus(list - set));
}
```

Figure 2: Pseudocode that illustrates how queries can be designed to harvest Web forms from a search engine.

Web in a focused way trying to find pages similar to a given description. The description could be a query that yields many pages with email-collecting forms.

An even more straightforward approach is for an agent to harvest forms from the Web by posting appropriate queries directly to some search engine. The agent can then fetch the hit pages to extract forms. For example MSN reports about 5 million hits for the query "free email newsletter" and over 800,000 hits for "send free SMS." However, search engines often do not return more than some maximum number of hits (say, 1,000). One way for the attacker's software to get around this obstacle is to create many query combinations by including positive and/or negative term requests. These combinations can be designed to yield large sets of hits with little overlap. Figure 2 illustrates how to create such queries automatically.

Once a potential page is identified, it must be parsed by the agent to extract form information. The page may actually not contain a form, or contain a form that cannot be used as a launch pad. A heuristic approach can be used to identify suitable forms. For example, there must be at least one text input field and either its name or its default value must match a string like "email." Such a heuristic identifies potential launch pad forms with high probability. In our experiments, using a search engine with queries as shown in Figure 2 leads to a form harvest rate of about 40%. In other words, the heuristic yields about 4 potential launch pad forms from each 10 search engine hits.

Once suitable Web form URLs are collected, they could be shared among attackers much like email address lists are exchanged among spammers. The harvest rate would then be 100%. It is easy to write software that parses the HTML code of a Web page and extracts form information. This consists of a URL for the form action, the method (GET/POST), and a set of input fields, each with a name, a type/domain, and possibly a default value. The form information can be stored in a database.

### 2.4 Phase II: Automatically Filling Forms

A form can be filled and submitted automatically, either immediately upon discovery, or at a later time based on the stored form's information. Heuristics can be used to

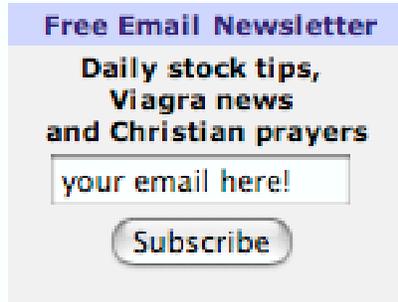

```
<form action="newsletter.php"
    method="POST">
<input type="text"
    name="Email"
    value="your email here!">
<input type="submit"
    name="submit"
    value="Subscribe">
</form>
```

Figure 1: A typical Web form that can be exploited by our attack (left), and the HTML code that can be used to detect, parse, and submit such a form (right).

assign values to the various input fields. These include the victim's email address and, optionally, other information such as name, phone, etc. Other text fields can be left blank or filled with junk. Fields that require a single value from a set (radio buttons, drop-down menus) can be filled with a random option. Fields that allow multiple values (checkboxes, lists) can be filled in with all options.

Once all input names have an associated value, an HTTP request can be assembled based on the form's method. Finally, sending the request for the action URL corresponds to submitting the filled form. For efficiency, forms can be filled and submitted in parallel by concurrent processes or threads.

This phase of the attack requires a form database, which could be a simple text file, and a small program that fills forms acting like a Web user agent (browser). The program could be executed from a public computer, for example in a library or a coffee shop. All that is required is an Internet connection. The program could be installed from a floppy disk, downloaded from a Web or FTP server, or even invoked via an applet or a virus.

### 2.5 Poorly Behaved Sites

There are many poorly behaved sites that may not care whether the entered contact information corresponds to the Web page visitor or a potential victim. The reason is simple: these sites derive benefit from the collection of valid email addresses, whatever their origin may be. The benefit may be the actual use of these addresses, or the sale of the same. For example, it is believed that the age verification scripts of many porn sites are simply disguised collectors of email addresses. We note that posting an email address to such a site may result in what we refer to as a *snow-balling effect*, i.e., a situation in which a submitted email address results in several emails, as the email address is bought, sold, and used.

The snow-ball effect can be exploited to maximize damage by generating a large-volume, persistent stream of email toward the victim. An efficient approach to maximize the number of spammers who obtain the victim's email is to post it on newsgroups and chatrooms, which are regularly and automatically scanned by spammers to harvest fresh email addresses. This approach does not even require one to collect and fill Web forms; but it has a more delayed, long-term effect.

### 2.6 Well Behaved Sites

While it is evident that the vulnerability we describe is made worse if the launch pads of the attack are poorly behaved sites, we argue that an attacker also can take advantage of well behaved sites. These are sites that may not sell the email address entered in the form, and who may wish to verify that it corresponds to a legitimate request for information. However, as previously mentioned, this typically involves sending an email to the address entered in the form, requesting an acknowledgement before more information is sent. This email, while perhaps not as large as the actual requested information, also becomes part of the attack as confirmation messages flood the victim's mailbox.

Moreover, if the intention of the form is to allow a user to send information to a friend, the above measures of caution are not taken. Examples of sites allowing such requests are electronic postcard services, many online newspapers, and more.

An attacker may also pose as a buyer to an e-commerce site, entering the victim's email address along with other information, such as an address and potentially incorrect credit card information. This would cause one or more emails to be sent to the victim. Given that the victim would not likely respond to any of these, the company may attempt to call the phone number entered in the form, which would constitute a potential attack in itself.

## 2.7 On the Difficulty of Tracing an Attacker

As described, the attack consists of two phases: one in which suitable forms are harvested and a second in which the forms are filled and submitted. While it is possible for a site to determine the IP address of a user filling a form, not all sites may have the apparatus in place to do so. Moreover, given the very short duration of the second phase (see section 3), it is easy for an attacker to perform this part of the attack using a public machine as shown above.

While the first phase of the attack typically takes more time, this can be performed once for a large number of consecutive attacks. Even if the first phase of the attack takes place from an identifiable computer and using a search engine, it is difficult for the search engine to recognize the intent of an attacker from the queries, especially considering the large numbers of queries handled. And it is impossible for a launch pad site to determine how its form was found by the attacker, whether a search engine was used, which one, and in response to what query. In other words, the second phase of the attack cannot be traced to the first (possibly traceable) phase.

Finally, the possibility of an attack — or parts thereof — being mounted by a virus (and therefore, from the machine of an innocent person) further frustrates any remaining hopes of meaningful traces.

## 3 Experimental Data

### 3.1 Experimental Setup

Here we report on a number of contained experiments carried out to demonstrate the ease of mounting the attack and its potential damage. We focus on email (as opposed to SMS) attacks in these experiments. We are interested in how many email messages, and how much data, can be targeted to a victim's mailbox as a function of time since the start of an attack. We also want to measure how long it would take to disable a typical email account.

Clearly these measurements, and the time taken to mount an attack, depend on the number of forms used. It would not be too difficult to mount an attack with, say, $10^5$ or $10^6$ forms. However, much smaller attacks suffice to disable a typical email account by filling its inbox. Furthermore, experimenting with truly large-scale attacks would present ethical and legal issues that we do not want to raise. Therefore we limit our experiments to very contained attacks, aiming to observe how the potency of an attack scales with its computational and storage resource requirements. We created a number of temporary email accounts and used them as targets of attacks of different sizes. Each attack used a different number of Web forms, sampled randomly from a collection of about 4,000 launch pads, previously collected.

In the collection phase of the attack, we used a "form-sniffing" agent to search the Web for appropriate forms based on hits from a search engine, using the technique described in section 2. The MSN search engine was used because it does not disallow crawling agents via the robot exclusion standard.[3] This was done only once.

The collection agent was implemented as a Perl script using no particular optimizations (e.g., no timeouts) and employing off-the-shelf modules for Berkeley database storage, HTML parsing, and the LWP library for HTTP. The agent crawled approximately 110 hit pages per minute, running on a 466 MHz PowerMac G4 with a 100 Mbps Internet connection. This configuration is not unlike what would be available at a copy store. From our sample we measured a harvest rate of 40% (i.e. 40 launch pad forms per 100 search engine hits) with a standard error of 3.5%. At this harvest rate, the agent collected almost 50 launch pad forms per minute, and almost 4,000 forms in less than 1.5 hours. If run in the background (e.g., in the form of a virus), this would produce as many as 72,000 forms in one day, or a million forms in two weeks — probably in significantly less time with some simple optimizations.

The second phase, repeated for attacks of different size, was carried out using the same machinery and similarly implemented code. A "form-filling" agent took a victim's information (email and name) as input, sampled forms from the database, and submitted the filled forms. The agent filled approximately 116 forms per minute. We call *attack time* the time required to mount an attack with a given number of forms.

### 3.2 Results

Figures 3 and 4 illustrate how the number of messages in the victim's inbox and the inbox size, respectively, grow over time after the attack is mounted. The plots highlight two distinct dynamic phases. While the attack is taking place, some fraction of the launch pad forms generate immediate messages toward the target. These responses correspond to an initial high growth rate. Shortly after the attack is over, the initial responses cease and a second phase begins in which messages continue to arrive at a lower, constant rate. These are messages that are sent by launch pads at regular intervals (e.g., daily newsletters), repeat acknowledgment requests, and spam. In the plots, we fit this dynamic behavior to the model

$$M_F(t) = (a_F \cdot t + b_F) \tanh(c_F \cdot t) \quad (1)$$

---

[3]We wanted to preserve the ethical behavior of the agent used in our experiments; an actual attacker could use any search engine since the robot exclusion standard is not enforceable.

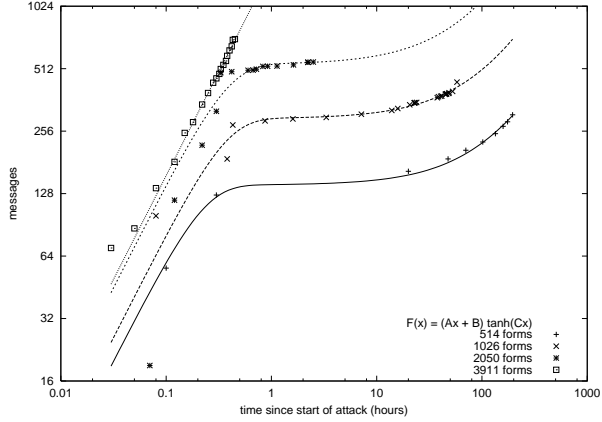

Figure 3: Number of messages received by victim versus time for attacks of different size.

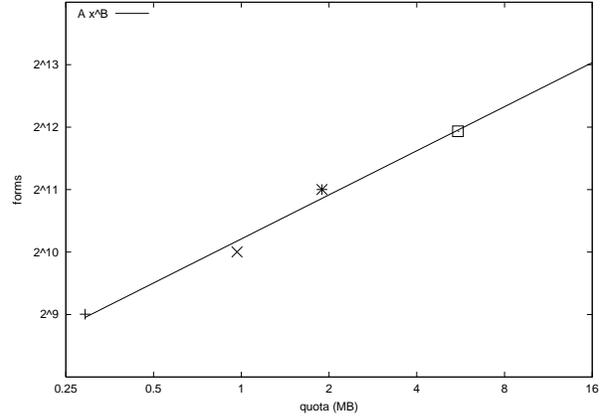

Figure 5: Attack size necessary to kill an account in an hour versus victim's quota.

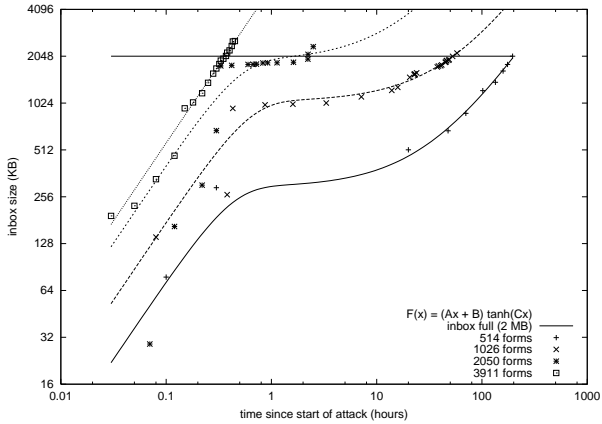

Figure 4: Victim's inbox storage versus time for attacks of different size. The account is killed when the inbox reaches 2 MB.

where $M_F(t)$ is the inbox size or number of messages at time $t$ ($t = 0$ is the start of the attack), for an attack with $F$ forms. The constants $a_F, b_F, c_F$ are determined by a nonlinear least-squares fit of the model to the data. The linear part of the model corresponds to the long-term growth, and $a_F$ is the stable arrival rate once the immediate responses have subsided, after the end of the attack. The hyperbolic tangent is a simple way to model the faster initial arrival rate. The initial phase is over when $\tanh(c_F \cdot t \gg 0) \approx 1$.

The email traffic generated by our attacks was monitored until the size of the inbox passed a threshold of 2 MB. This is a typical quota on free email accounts such as Hotmail and Yahoo. No other mail was sent to the victim accounts, and no mail was deleted during the experiments. When an inbox is full, further email is bounced back to senders and, for all practical purposes, the email account is rendered useless unless the victim makes a significant effort to delete messages. We call *kill time* the time between the start of an attack and the point when the inbox size reaches 2 MB.

In Figures 3 and 4 we can observe that for the three smaller attacks ($F = 514, 1026, 2050$) kill time occurs well after the attack has terminated. For the largest attack ($F = 3911$), kill time occurs while the attack is still being mounted. This is mirrored by the fact that this attack is still in the initial phase of high response rate when the inbox fills up.

One can use the data of Figure 4 and the model of Equation 1 to analyze how large an attack would be necessary to kill an account in a given amount of time, as a function of the account quota. Figure 5 shows the number of forms that in our experiments would kill an account in one hour, corresponding to a *lunch hour attack*, in which the victim's machine is disabled while she is temporarily away. The number of forms scales sub-linearly, as a power law $F \sim q^{0.7}$ where $q$ is the account quota. We can think of this as a manifestation of the snow-ball effect — periodic alerts and spam compound immediate responses making the attack more efficient.

Figure 6 shows how the arrival rate of email in the victim's mailbox scales with the size of the attack. The arrival rate for an attack of size $F$ is given by the growth parameter $a_F$, obtained by fitting the model in Equation 1 to the data in Figure 4. As illustrated by the least-squared fit in Figure 6, the arrival rate appears to scale exponentially: $a_F \sim e^{0.0019F}$. Such a non-linear scaling behavior is surprising; it is the result of the fact that $a_{3911}$ is the short-term (not long-term) growth rate, as the attack is still being mounted. If we only consider the long-term growth rate and assume there is no snow-ball effect, the arrival rate should scale linearly with $F$. To illustrate this, Figure 6 also plots a linear fit of the arrival rate data for the small attacks: $a_F \sim 0.06F$ for $F \leq 2050$.

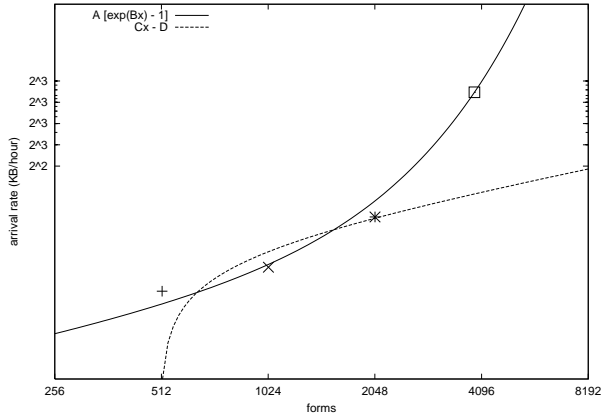

Figure 6: Growth rate of victim's inbox versus attack size.

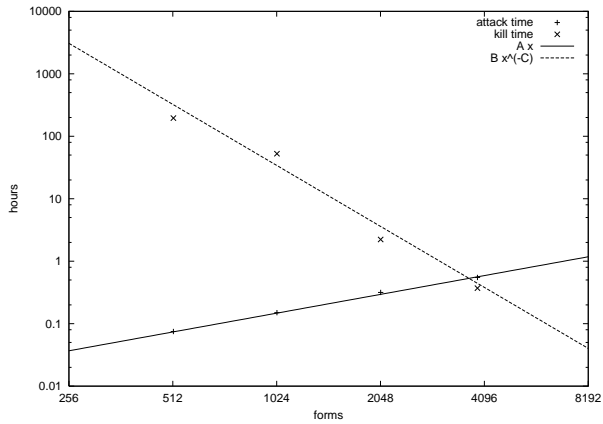

Figure 7: Attack time and time to fill a 2 MB inbox, as a function of attack size.

Finally, Figure 7 shows how attack time and kill time scale with the size of the attack. As expected attack time is proportional to $F$. Kill time (cf. Figure 4) scales as a power law: $t \sim F^{-3.2}$. Again, this non-linear scaling behavior is a consequence of the snow-ball effect, which amplifies the destructive effect of the attack and makes it possible to kill an email account efficiently. In fact the intersection between attack and kill time in Figure 7 indicates that there is no need to mount attacks with more than $F \approx 2^{12}$ forms if the goal is to disable an account with a 2 MB quota.

## 4 Survey of Vulnerable Targets

**Attacking known targets.** First of all, it is clear that an attacker could simply target users with email addresses known to him, including addresses that result in a text message being sent to a cellular phone or other mobile device. Many people make their email addresses known to the public, whether in machine readable format or not; many companies rely on making public addresses to costumer service (making them susceptible to attacks both from competitors and people who disagree with their policies or merchandise); and oftentimes, politicians make both their phone numbers and email addresses available to the public. (While these certainly are not their home phone numbers or email addresses given to fellow politicians, they are important means for their constituents to reach them.) Some eBay users appear to use their email addresses as identifiers, making it easy to block these from any competition during an auction.[4] A large portion of the remaining set of eBay users can be conned into giving out their email address: simply ask them an innocuous question relating to a previous transaction of theirs (using the supplied Web interface) and the reply will contain their email address. Furthermore, many banks use email for internal and external communication, opening up for an attack on their computers. Many journalists and law enforcement officers rely on phone and email for leads and pointers.

**Attacking random targets.** Given that many companies use highly predictable formatting for email addresses, it may be possible for an attacker to mount an attack on people believed to work for the company, or on people with common names, which in the end may amount to an attack of the company itself. Furthermore, an attacker can use the very same tools spammers use to harvest and purchase valid email addresses, filtering these to suit the profile of his attack. People can be attacked based on their likely geographical location by selecting phone numbers with given area codes. Even if a rather small percentage of randomly selected phone numbers correspond to actual cell phones with text messaging capabilities, the price of mounting an attack is so low that an attacker may try random numbers. If a phone-based attack is mounted, an attacker may use on-line phone books to select victims, whether the selection of the victims is automated or manual, and whether it is targeted or random.

**Politically motivated attacks.** If a large number of random mobile devices are attacked during an electronic election, it is highly probable that some voters will be unable to cast their vote. While this is not likely to swing the election results, unless the attack is severe, it will cause the results, and the fairness of these, to be questioned. This may especially be so if the targeted phone numbers correspond to particularly rich or poor voting districts, or to districts with higher proportions of certain minorities. Apart from attacking the political infrastructure (which may cause disruption, particularly during elections), an attacker may attempt to target people of certain ethnicities or nationalities by performing filtering based on particular names or affiliations. Attackers may target people based on their likely opinions or inclinations by harvesting email addresses from selected bulletin boards, or by performing a focused crawl for given keywords on personal Web pages. Attackers may attempt to bring down selected email-based chatrooms by generating traffic to them — notice that it does not matter whether they are moderated or not, since a moderator cannot approve good posts if the bad ones are just too overwhelming.

---

[4]This attack succeeds unless the victim has already entered a sufficiently high maximum bid, allowing eBay to act on their behalf. Still, though, if a winning bidder's email account is successfully disabled before the payment has been performed, the victim will not be able to respond to the seller, making the latter likely to contact the second-highest bidder or re-run the auction.

## 5 Defense Mechanisms

We now describe a set of related defense techniques for our DDoS attack. A first line of defense consists of a simple *preventive* step by which Web sites can avoid being exploited as launch pads in our attack. For Web sites that have not yet complied with this preventive step, as well as unscrupulous spammer sites that have no intention to verify the legitimacy of requests, we describe a second line of defense for the *detection* and *management* of such attacks by potential victims. The second line of defense consists of a heuristic approach, whose use can be adapted to different situations of interest; we focus on three typical entities that differ in the types of emails they are likely to receive: an individual, an on-line store, and a politician.

### 5.1 Prevention of Attacks

Many sites that allow users to subscribe to email services such as newsletters and alerts employ `mailto` links (either to a person or to a listserv manager, e.g., Majordomo). These sites cannot be exploited as launch pads, because the attacker would need a mail transport agent, e.g. a machine running a SMTP server or an external mail relay. Such an attack is possible, but more difficult to carry out from a public computer and also more easily detectable and traceable. Open relays are rare and often blocked by ISPs anyway (because they are used by spammers), and a "legitimate" SMTP server requires some level of authentication that would allow to identify or trace the attacker. The obvious preventive solution to the proposed attack is thus to disable Web forms and enforce the use of email-based listserv tools such as Majordomo. However, this would disallow useful Web forms in which users can enter additional information — this cannot be done conveniently with a simple `mailto` link to a listserv. To allow for the use of forms as appropriate while still verifying the legitimacy of email service requests, well behaved sites currently send a message to the submitted email address requesting confirmation that the address corresponds to a legitimate user request. As we observed earlier this behavior is exploited in our attack because confirmation requests, even if not repeated (as they often are), contribute to flooding the victim's mailbox just as any other message.

It is possible to both enable Web form requests and verify the legitimacy of requests, without becoming vulnerable to our attack. Web sites would use the following simple strategy. After the form has been filled out, the Web site creates dynamically a page containing a `mailto` link with itself as an addressee. Legitimate users would send the message to validate their request. The email to the Web site would then be used by the site's mailing list manager to verify that the sender matches the email address submitted via the Web form. Although the address of the sender is not reliable because it can be spoofed in the SMTP protocol, the sender cannot spoof the IP address of its legitimate ISP's SMTP server. The site can thus verify that the email address in the form request matches the originating SMTP server in the validation message.

There are three caveats to this strategy. First, messages via open relays must be discarded by the site. Second, if an attacker could guess that a user in a given domain requests information from some site, she could request information from the same site for other users in the same domain, potentially spoofing the validation created by the addressee. To prevent such an attack, the validation message created by the site should contain a number with sufficient entropy that it is hard to guess. Third, one could still attack victims who share their ISP's mail server. In general this would be somewhat suicidal, but a disgruntled employee might use such an attack against his employer. In this case, however, the attack could be traced. Furthermore, our heuristic defense mechanisms — presented next — will address such an attack. With these caveats, our preventive strategy would afford the same security as forms that now request email confirmation, but without sending any email to victims.

The above technique works for forms where a party requests information to be sent to herself, but it does not cover common services such as sending newspaper articles or postcards to others. Sites wishing to allow this can use alternative defenses. Namely, well behaved sites may make the harvesting of forms more difficult by not labeling forms using HTML, but rather, using small images. This would increase the effort of finding and filling the forms. Given the relative abundance of available forms, potential attackers are then likely to turn to other sites where no image analysis has to be performed to find and fill the form. Doing this has no impact on human users, except to a very small extent on the download time of the form. A more robust version of this defense would use an inverse Turing test or CAPTCHA (Completely Automatic Public Turing test to tell Computers and Humans Apart) [11, 10], a technique already employed by many sites to prevent agents from impersonating human users.

If legislation is brought in place that makes sites liable for any attacks mounted using their facilities [8], then even poorly behaved sites may wish to employ protective measures as those described above to avoid being the defendants in lawsuits by victims of the attack we describe.

## 5.2 Detection and Management of Attacks

In the previous subsection, we considered how well-behaved sites can protect themselves against being used as launch pads. Since it is not likely that all sites will comply with these protective measures, we also need to consider protection against poorly behaved and otherwise non-compliant sites. This protection will reside on the side of the potential victim, whether on his machine or on his mail server. Before detailing the defense mechanisms, let us introduce three tools that these will employ:

**Extended Address Book.** Most users maintain an address book in which they enter the email addresses of their most frequent correspondents. We consider the use of an additional address book, what we will refer to as the *extended address book.* This contains the email addresses of all parties the user has sent email to or received email from, along with a time stamp indicating when the last email was sent or received. To reduce the required storage, we may allow users to have entries automatically removed after their corresponding date stamp reaches a given age selected by the user. The extended address book is similar to the whitelists maintained by spam filters; the main difference is that it would only be used for filtering purposes when an attack is suspected, as described below. Emails of spammers might even be included. When deemed beneficial, a set of users may share one and the same extended address book.

**Attack Meter.** We will let the system estimate the probability that a given user is under attack at any given time. The parameters considered would be the amount of traffic to the user in relation to the normal amount of traffic to her, and relative to the traffic of other users; the proportion of emails arriving to the user (and her peers) that originate from users that are not in their extended address books; and the number of duplicate emails received by users handled by the mail server. The calibration of the estimation may be performed with a given threat situation in mind.

**Cleaner.** During a clean-up, a set of suspect emails are removed from the inbox of the user. Depending on the situation, it may be that all suspect emails are removed; all suspect emails of some certain minimum size; all suspect emails from (or not from) given domains; or some other, potentially customized selection of all suspect emails. The mail server may automatically respond to the sender, notifying them that their email was erased, potentially using a notification customized by the recipient.

**Defense of an individual.** When a user accesses his account, he would be shown the likely probability, according to the attack meter, that he is under attack. If the user indicates that he believes he is under attack, the mail server would automatically mark all emails that are from senders who are *not* in the extended address book as *suspect,* and proceed to perform a clean-up. This may also be induced by the system — without the request of the user — if the user is not available, an attack is judged to likely be under progress, and resources are scarce. These defenses could reside either on the user side, or on the side of the service provider, as is appropriate for wireless devices. If residing with the provider, the attack meter can also take the general attack situation in consideration when determining whether an individual is being attacked. We note that this solution also secures list moderators at the expense of not being able to receive messages from new posters during the time of an attack; note also that the risk of the launch pads already being in the extended address book of the moderator is slim.

If a person wants to always be able to receive high-priority messages, then he can download email from multiple sources (e.g., using POP), and use an obscure address for the high-priority email. This address would only be known by the senders of high-priority messages, and would have sufficient entropy to make a dictionary attack improbable to succeed. If the user believes he is under attack, he can switch from synchronizing his device with all ISPs he uses to only the ISP of the high-priority account. This provides an increased level of protection against attacks for people on call, such as technical support staff, medical doctors, and more.

**Defense of an online store.** If considered under attack, the mail server would mark emails as suspect if they originate from a user who is not in the extended address book, unless this user is a known collector of email from other sources. For an example of the latter, consider how eBay users establish communication with each other: via the messaging interface of eBay. Thus, eBay serves the role of an email collector, and associated emails would be handled using particular rules. For example, the mail server may mark a set of emails as suspect if arriving in large quantities from one and the same user pseudonym; if associated with a newly established user pseudonym; or with a user pseudonym with limited or low feedback. After this, the clean-up is started.

**Defense of a politician.** A politician may have set up an email account to enable communication with his constituents. Many of these are likely to use accounts with one of a very small set of known ISPs. In contrast, companies responding to forms are likely not to have the same domains. Therefore, under attack, the mail server could mark as suspect those emails that do not come from the known ISPs likely to correspond to the wanted senders. Furthermore, the mail server may mark emails as suspects if coming from other countries — when indicated by the corresponding domain — as these are also unlikely to be from constituents.

## 5.3 Synergy between Defense of Launch Pads and Victims

It is important that the heuristic defense mechanisms proposed do not disrupt desired functionality, thus it must still be possible for a user to fill forms and receive information sent to him. Indeed this will still be possible — even during a detected attack — as long as the site with the form sends email from an address that is present in the extended address book of the party requesting information.

In the strategy described above to prevent Web sites from being exploited as launch pads, the user who submits a request through a form must send a validation message (dynamically created and self-addressed) to the Web site. This step causes the Web site's email address to be entered into the user's extended address book. As a result the information sent to the user by the site is not filtered out. This creates an incentive for sites to comply with the preventive strategy, not only to avoid being exploited but also to keep their messages from being filtered out.

## 6 Discussion

We investigated an automated and agent-based DDoS attack in which a victim is swamped by communication from entities believing she requested information. The primary tool of the attack is that of Web forms, which can be automatically harvested and filled out by an agent.

The automatic recognition and extraction of forms from Web pages using simple heuristics is not a new concept. For example it has been applied to the design of comparison shopping agents aimed at searching for products from multiple vendor sites [4]. The problem is only a bit harder if an account must be set up before a a form can be submitted. For instance many sites allow only registered users to send SMSs to any number. However, setting up an account is free and can easily be automated — this is why, e.g., Hotmail and Yahoo use CAPTCHAs to prevent spammers from setting up fake accounts automatically.[5]

During a denial of service attack a large number of connections is set up with a victim, thereby exhausting the resources of the latter. A distributed denial of service attack is mounted from multiple directions, thereby making it more difficult to defend against. There exist many automated tools to mount DDoS attacks [2, 3, 5]. These require that the attacker takes control of a set of computers from which he will launch the attack. This, in turn, makes DDoS attacks more difficult to perform for a large portion of potential offenders. It also offers a certain degree of traceability since the take-over of launch pad computers may set off an alarm. The poor man's DDoS attack illustrated here can be mounted without the need to take over any launch pad computer, and offers the offender an almost certain guarantee of untraceability — due both to its swiftness and to the fact that it utilizes only steps that are also performed by benevolent users.

We described a novel, very simple strategy by which Web sites can avoid being exploited in the poor man's DDoS attack; once a majority of Web sites comply with this strategy, such attacks will be prevented.

For the interim, we have proposed a set of heuristic techniques to inoculate users against the poor man's DDoS attack. These mechanisms only allow emails to be filtered out if they are sent from sites that are not in a user's extended address book. While there is no cryptographic mechanism to avoid IP spoofing, this is not a major threat because it is not the attacker who would have to spoof the IP address of the sender of the email, but the launch pad site. Well behaved sites will clearly not do this, and if poorly behaved sites are willing to, then they become part of the aggressor, and not merely a tool in its hand. What makes our attack severe is that the launch pads would be oblivious to the role they are playing, and that is not the case if they perform IP spoofing. If a site is willing to spoof IP addresses, there are clearly simpler DoS attacks such as mailbombs that do not involve Web forms or agents.

Our attack is an extension and variant of the recent work by Byers, Rubin and Kormann [1], in which an attack was described where victims are inundated by *physical* mail. While the underlying principles are the same, the ways the attacks are performed, and what they achieve, are different. By generalizing to mostly all types of communication, our attack becomes a weapon in the hands of an attacker wishing to attack secondary targets as well as primary ones. Moreover, the defenses proposed in the two papers vary considerably, given both the different threat situations and the different goals in terms of systems to be secured. We consider both how to secure entities against becoming victims and how to secure sites against being exploited as launch pads, while the work of [1] only considers the latter. This strengthens our defenses in the face of poorly behaved Web sites, and non-compliant sites.

---

[5]Incidentally, it would be beneficial for eBay to do so as well during the account creation phase, or their service remains vulnerable against an agent based attack in which a large number of accounts are created and later used for disruptive purposes. One such disruptive purpose would be to bid up on items — without later paying for them — thereby blocking legitimate bidding.

We have not investigated the generation of traffic by means of posting messages to newsgroups, chatrooms and bulletin boards, purportedly from the victim, but believe such attacks to be similar to those we discussed, and possible to defend against in similar manners.

There are more drastic types of defense measures that can protect from the attack described in this paper. For example some ISPs are considering CAPTCHA based challenge-response systems in conjunction with whitelists to combat spam.[6] While such an approach would indeed protect a potential victim from the email DDoS attack, it would also decrease the accessibility of email. Many email-based transactions, such as e-commerce confirmations, would also be blocked. The defenses we have described are more targeted at the DDoS attack, more lightweight, and do not require modifications to the email infrastructure.

At a more general level, the kind of attack described here raises new issues with social, ethical, legal and political implications for the use and integration of modern communication media such as the Internet, electronic messaging, and mobile telephony. For example, if users were required to identify themselves when using the Internet in order to prevent such abuses, then one could no longer use a computer anonymously in a public place such as a library. We hope that this work will spark a fruitful debate on these issues, leading to solutions that will protect our inboxes as well as our privacy and freedom of speech.

## Acknowledgments

We thank Avi Rubin, Aleta Ricciardi, John Linn, Burt Kaliski and Shannon Bradshaw for useful discussions. This work was supported in part by NSF Career Grant No. IIS-0133124 to FM.

## References

[1] S. Byers, A. Rubin, and D. Kormann. Defending against an Internet-based attack on the physical world. In *Proc. ACM Workshop on Privacy in the Electronic Society*, 2002.

[2] S. Dietrich, N. Long, and D. Dittrich. Analyzing distributed denial of service tools: The Shaft case. In *Proc. 14th Systems Administration Conference (LISA 2000)*, 2000. http://www.usenix.org/events/lisa2000/dietrich.html.

[3] D. Dittrich. Distributed denial of service (DDoS) attacks/tools. http://staff.washington.edu/dittrich/misc/ddos/, 2003.

[4] R. Doorenbos, O. Etzioni, and D. Weld. A scalable comparison-shopping agent for the World-Wide Web. In *Proceedings of the First International Conference on Autonomous Agents*, pages 39–48, 1997.

[5] K. Houle, G. Weaver, N. Long, and R. Thomas. Trends in denial of service attack technology. CERT Coordination Center White Paper, October 2001. http://www.cert.org/archive/pdf/DoS_trends.pdf.

[6] F. Menczer, G. Pant, M. Ruiz, and P. Srinivasan. Evaluating topic-driven Web crawlers. In D. H. Kraft, W. B. Croft, D. J. Harper, and J. Zobel, editors, *Proc. 24th Annual Intl. ACM SIGIR Conf. on Research and Development in Information Retrieval*, pages 241–249, New York, NY, 2001. ACM Press.

[7] F. Menczer, G. Pant, and P. Srinivasan. Topical web crawlers: Evaluating adaptive algorithms. *ACM Transactions on Internet Technology*, Forthcoming, 2003. http://dollar.biz.uiowa.edu/~fil/Papers/TOIT.pdf.

[8] J. Silva. Spam small problem ... today. RCRNews, 2003. http://www.rcrnews.com/cgi-bin/article.pl?articleId=42294.

[9] SkyNews. Elections: The final push. http://www.sky.com/skynews/article/0,,30100-12300859,00.html, 2003.

[10] L. von Ahn, M. Blum, N. Hopper, and J. Langford. CAPTCHA: Using hard AI problems for security. In *Proceedings of Eurocrypt*, 2003.

[11] L. von Ahn, M. Blum, and J. Langford. Telling humans and computers apart (automatically). *Communications of the ACM*, forthcoming.

---

[6]Earthlink has announced a beta version of such a system as of this writing.